\documentclass{article}
\usepackage{spconf,amsmath,graphicx,subfigure,multirow}

\title{Deep denoising auto-encoder for statistical speech synthesis}
\twoauthors
{Zhenzhou Wu\sthanks{The first author performed this work while at the National Institute of Informatics, Japan.}}
{School of Computer Science \\
McGill University, Canada}
{Shinji Takaki, Junichi Yamagishi\thanks{This work was supported in part by EPSRC through Programme Grant EP/I031022/1 (NST) and EP/J002526/1 (CAF). Shinji Takaki was supported in part by NAVER Labs.}}
{National Institute of Informatics \\
Japan}
\begin{document}

\ninept
\maketitle

\begin{abstract}
%%%%% first submitting version %%%%%
%In this paper we propose a deep denoising auto-encoder technique to extract better acoustic features for speech synthesis. This allows us to automatically %extract low dimensional features from high dimensional spectral features in a non-linear, data-driven, unsupervised way. We compared the new stochastic feature %extractor with the conventional mel-cepstral analysis in analysis-by-synthesis and text-to-speech experiments. Our results confirm that the proposed method can %increase the quality of synthetic speech in both experiments.
This paper proposes a deep denoising auto-encoder technique to extract better acoustic features for speech synthesis. The technique allows us to automatically extract low-dimensional features from high dimensional spectral features in a non-linear, data-driven, unsupervised way. We compared the new stochastic feature extractor with conventional mel-cepstral analysis in analysis-by-synthesis and text-to-speech experiments. Our results confirm that the proposed method increases the quality of synthetic speech in both experiments.

\end{abstract}
\section{Introduction}
\label{sec:intro}

Current statistical parametric speech synthesis typically uses hidden Markov models (HMMs) to represent probability densities of speech trajectories given text \cite{ref:Zen09}. This is a well-established method and it is straightforward to apply this framework for new languages. It also offers interesting advantages in terms of flexibility and compact footprint \cite{ref:Yoshimura97,ref:Yoshimura99,ref:Tsuzuki04,ref:Yamagishi05}. It is known, however, that speech synthesized from statistical models still sounds somehow artificial and less natural compared to speech synthesized by the best unit selection systems.

It is often said that averaging in statistical synthesis systems partly removes spectral fine structure of natural speech, and thus there is room for the improving the segmental quality. A stochastic postfilter approach \cite{ref:Chen14} proposes to use a deep neural network (DNN) to model the conditional probability of the spectral differences between natural and synthetic speech. The approach is able to reconstruct the spectral fine structure lost during modeling and has achieved significantly quality improvement for synthetic speech \cite{ref:Chen14}. In this experiment, the HMM-based speech synthesiser was trained in the mel-cepstral domain, while the DNN-based postfiler was trained in the spectral domain.

This indicates that the current statistical parametric speech synthesis suffers from quality loss due to statistical averaging in the mel-cepstral domain, but also due to conversion from high-dimensional spectral features to lower dimensional mel-cepstral parameters and hence this brings us a new question: are current intermediate representations such as mel-cepstral coefficients appropriate for statistical training of acoustic models? Can we automatically find a more appropriate intermediate representation that suits acoustic modelling and results in better quality of synthetic speech?

To answer this question, this paper proposes a DNN-based feature extraction method. More specifically we propose to use a deep denoising auto-encoder technique as a non-linear robust feature extractor for speech synthesis and apply it to high-dimensional spectral features obtained from STRAIGHT vocoder \cite{ref:Kawahara99}. We compare this data-driven, unsupervised feature extraction approach with the conventional mel-cepstral analysis, which is based on a linear discrete cosine transform of the log spectrum.

This paper is organised as follows: in Section 2, we outline related DNN-based approaches and in Section 3 we describe the proposed deep denoising auto-encoder technique. In Section 4, we mention how we train the model and the experimental conditions and evaluation results are shown in Section 5. Discussions and the summary of our findings are given in Section 6.

\section{Related work using DNN and auto-encoder}
\label{sec:related_work}

This section overviews related work using DNN and/or auto-encoder in the speech information processing field. DNN has been applied for acoustic modelling of speech synthesis. For instance \cite{ref:Zen13} uses DNN to learn the relationship between input texts and the extract features instead of decision tree-based state tying. Restricted Boltzmann machines or deep belief networks have been used for modelling output probabilities of HMM states instead of GMMs \cite{ref:Ling14}. Recurrent neural network or long-short term memory was used for prosody modelling \cite{ref:Fan14} or acoustic trajectory modelling \cite{ref:Fernandez14}.

To the best of our knowledge, this is the first work to use deep denoising auto-encoder for speech synthesis, but, deep auto-encoder based bottleneck features are used by several groups for ASR \cite{ref:Sainath12,ref:Gehring13} and deep denoising auto-encoder is also verified for noise-robust ASR \cite{ref:Maas12} or reverberant ASR tasks \cite{ref:Ishii13,ref:Feng14}.

Techniques that are closely related to this paper are a spectral binary coding approach using deep auto-encoder proposed by Deng et al \cite{ref:Deng10} and a speech enhancement approach using deep denoising auto-encoder where they try to reconstruct clean spectrum from noisy spectrum \cite{ref:Lu13}. The approach proposed here is also related to heteroscedastic linear discriminant analysis (HLDA) \cite{ref:Kumar98,ref:Gales02} and probabilistic linear discriminant analysis (PLDA) \cite{ref:Prince07,ref:Kenny10,ref:Lu14}. Our key idea is however different from these as we use deep auto-encoder based continuous bottleneck features calculated from spectrum to reconstruct high-quality synthetic speech.

\section{Auto-encoder}
\label{sec:auto-encoder}

\subsection{Basic Auto-encoder}
\label{ssec:basic_auto-encoder}

Auto-encoder is an artificial neural network that is used generally for learning a compressed and distributed representation of a dataset. It consists of the encoder and the decoder. The encoder maps a input vector $\mathbf{x}$ to a hidden representation $\mathbf{y}$ as follows:
\begin{equation}
\mathbf{y} = f_\theta(\mathbf{x}) = s(\mathbf{Wx + b}),
\end{equation}
where $\theta = \{\mathbf{W}, \mathbf{b}\}$. $\mathbf{W}$ and $\mathbf{b}$ represent a $m \times n$ weight matrix and a bias vector of dimensionality $m$ respectively, where $n$ is the dimension of $\mathbf{x}$.
The function $s$ is a non-linear transformation on the linear mapping $\mathbf{Wx+b}$. Frequently $s$ is a sigmoid, a tanh, and a relu  function.
$\mathbf{y}$, the output of the encoder, is then mapped to $\mathbf{z}$, the output of the decoder.
The mapping is performed by a linear function alone that employs a $n \times m$ weight matrix $W'$ and a bias vector of dimensionality $n$ as follows:
\begin{equation}
\mathbf{z} = g_{\theta'}(\mathbf{y}) = \mathbf{W'y + b'},
\end{equation}
or a linear mapping followed by a non-linear transformation $t$,
\begin{equation}
\mathbf{z} = g_{\theta'}(\mathbf{y}) = t(\mathbf{W'y + b'}),
\end{equation}
where $\theta' = \{\mathbf{W', b'}\}$. The weight for the decoding is set as the transpose of the encoding weight \cite{ref:Hinton06} in order to allow more layers to be stacked together and be fine-tuned with stochastic gradient descend (SGD).

In general, the output $\mathbf{z}$ should be interpreted as a function of parameters $\{\theta,\theta'\}$ as $\mathbf{z} = g_{\theta'}(f_{\theta}(x))$. The parameters $\{\theta, \theta'\}$ are optimized such that the reconstructed $\mathbf{z}$ is as close as possible to the original $\mathbf{x}$ and maximizes $P(\mathbf{x|z})$. A typical loss function used is the mean square error (MSE), i.e. $L(\mathbf{x,z}) = \frac{1}{n}|\mathbf{x-z}|^2$.

\subsection{Denoising Auto-encoder}
\label{sssec:subsubhead}

The denoising auto-encoder is a variant of the basic auto-encoder. It is reported that the denoising auto-encoder can extract features more robustly than the basic auto-encoder \cite{ref:Vincent08}. In the denoising auto-encoder, the original data $\mathbf{x}$ is first corrupted to $\mathbf{\tilde{x}}$ before it is mapped to a higher representation $f_{\theta}(\mathbf{\tilde{x}})$ by an encoder.
The decoder then maps the higher representation to the output $\mathbf{z}$ for reconstructing the original $\mathbf{x}$.
The denoising auto-encoder is trained such that the reconstructed ${\mathbf{z}}$ is as close as possible to the original data ${\mathbf{x}}$. Note that it is only during training that the denoising auto-encoder is used to reconstruct the original $\mathbf{x}$ from the corrupted $\mathbf{\tilde{x}}$.

%In ASR, the denoising auto-encoder is frequently used to extract noise-robust features.
%\textbf{Differing from ASR, in speech synthesis, the reconstruction from bottleneck features is a key aspect, the input and reconstructed spectra have much higher dimension and over-training problem may occur. The denoising auto-encoder would alleviate this problem and extract more appropriate features.}
%\textbf{I think this is not clear - zhenzhou}

\subsection{Deep Auto-encoder}
Auto-encoder or denoisng auto-encoder can be made deeper by stacking multiple layers of encoders and decoders to form a deep architecture. \cite{ref:Yu11} shows that deeper architecture produces better high-level features compared to the shallow architecture up to 4 encoding and 4 decoding layers. For constructing a deep auto-encoder pre-training is widely used.
In pre-training, the number of layers in a deep auto-encoder increases twice as compare to a deep neural network (DNN) when stacking each pre-trained unit. It is reported that fine-tuning with back-propagation through a deep auto-encoder is ineffective due to vanishing gradients at the lower layers \cite{ref:Sepp01}. To over come this issue we restrict the decoding weight as the transpose of the encoding weight following \cite{ref:Hinton06}, that is, $\mathbf{W}'=\mathbf{W}^T$ where $\mathbf{W}^T$ denotes transpose of $\mathbf{W}$. We describe the detail of training a deep auto-encoder in the next session.

\section{Training a Deep Denoising Auto-encoder}
\subsection{Greedy Layer-wise Pre-training}
\begin{figure}[t]
  \begin{center}
  \includegraphics[width=8.0cm]{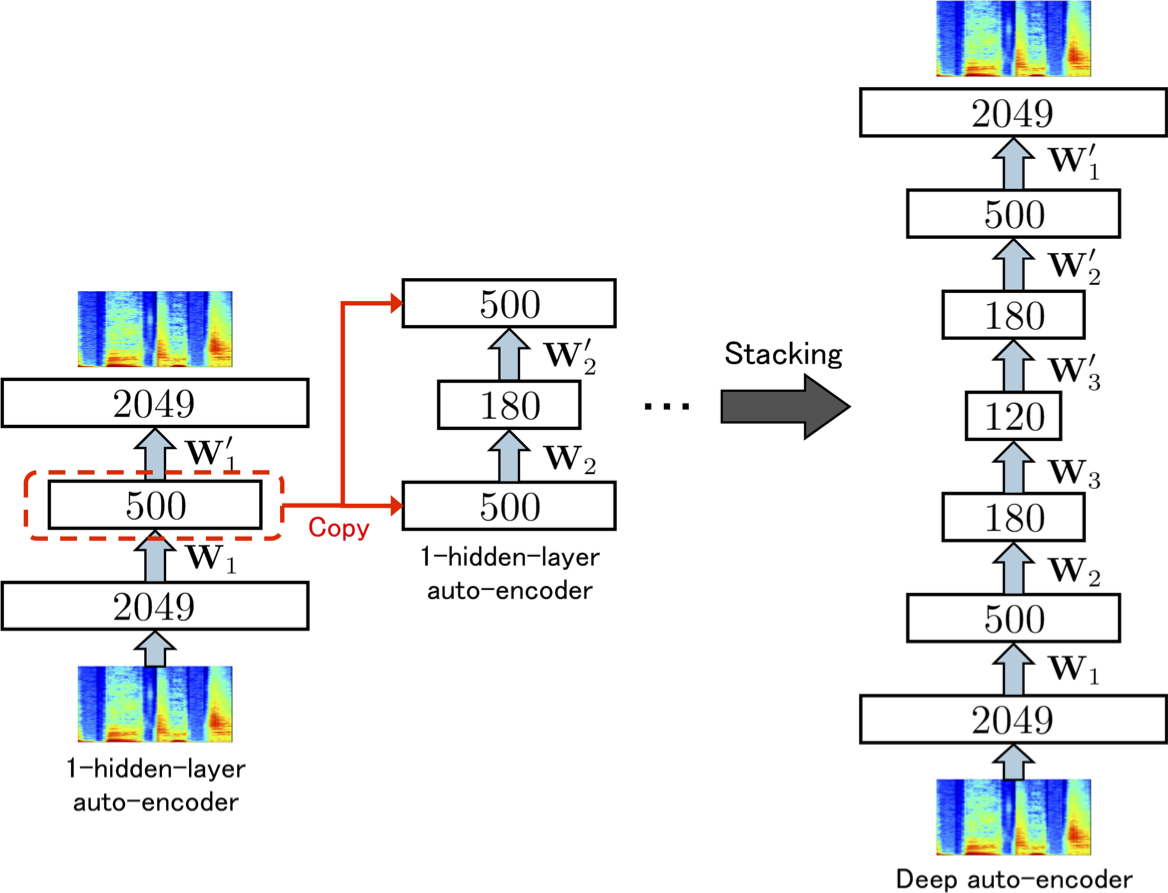}
  \end{center}
\vspace{-0.5cm}
\caption{Greedy layer-wise pre-training for constructing deep auto-dencoder.}
\label{fig:Stacking_AEs}
\vspace{-5mm}
\end{figure}
Each layer of a deep auto-encoder can be pre-trained greedily to minimize the reconstruction loss $L(\mathbf{x},\mathbf{z})$ of the data locally. Figure \ref{fig:Stacking_AEs} shows a procedure of constructing a deep auto-encoder using pre-training.
In pre-training a 1-hidden-layer auto-encoder is trained and the encoding output of the locally trained layer is used as the input for the next layer.
This layer-wise training is repeated until the desired layer size is obtained.
The encoding, decoding and loss functions of each layer are represented as follows:
\begin{equation}
\begin{split}
\texttt{Layer 1:} \\
\mathbf{y}_1 &= f_{\mathbf{W}_1,\mathbf{b}_1}(\mathbf{x}), \\
\mathbf{z}_1 &= g_{\mathbf{W}'_1,\mathbf{b}'_1} (\mathbf{y}_1), \\
L(\mathbf{x},\mathbf{z}_1) &= |\mathbf{x} - \mathbf{z}_1|^2, \\
%\texttt{Layer 2:} \\
%\mathbf{y}_2 &= f_{\mathbf{W}_2,\mathbf{b}_2}(\mathbf{y}_1), \\
%\mathbf{z}_2 &= g_{\mathbf{W}'_2,\mathbf{b}'_2}(\mathbf{y}_2), \\
%L(\mathbf{y}_1,\mathbf{z}_2) &= |\mathbf{y}_1 - \mathbf{z}_2|^2, \\
%\vdots \\
\texttt{Layer k (k>1):} \\
\mathbf{y}_k &= f_{\mathbf{W}_k,\mathbf{b}_k}(\mathbf{y}_{k-1}), \\
\mathbf{z}_k &= g_{\mathbf{W}'_k,\mathbf{b}'_k}(\mathbf{y}_k), \\
L(\mathbf{y}_{k-1},\mathbf{z}_k) &= |\mathbf{y}_{k-1} - \mathbf{z}_k|^2. \\
\end{split}
\end{equation}
Note that during the pre-training of the deep denoising auto-encoder, the input $\mathbf{x}$, $\mathbf{y}_k$ for each layer are corrupted to $\tilde{\mathbf{x}}$ and $\tilde{\mathbf{y}_k}$ respectively.
After all layers are pre-trained, all the pre-trained layers are stacked for constructing a deep denoising auto-encoder in the same way as the deep auto-encoder.

\subsection{Fine-tuning}
The purpose of fine-tuning is to minimize the reconstruction error $L(\mathbf{x,z})$ over the entire dataset and a model architecture using error back-propagation \cite{ref:Remelhart86}.
We use the mean square error (MSE) for the loss function of a deep auto-encoder and it is represented as follows:
\begin{equation}
E = \sum_{i=1}^N |\mathbf{x}^{(i)} - \mathbf{z}^{(i)}|^2,
\end{equation}
where $N$ is the total number of training examples.
The partial derivatives w.r.t weight $w_{i,j}^{(l)}$ is represented as follows:

\begin{equation}
\frac{\partial{E}}{\partial{w_{i,j}^{(l)}}} = \frac{\partial{E}}{\partial{t_j^{(l)}}} \times \frac{\partial{t_j^{(l)}}}{\partial{w_{i,j}^{(l)}}}
\end{equation}
where $t_j^{(l)}$ is the fan-in input to neuron $j$ in layer $l$, and $\frac{\partial{t_j^{(l)}}}{\partial{w_{i,j}^{(l)}}} = o_{i}^{(l-1)}$, where $o_{i}^{(l-1)}$ is the output from neuron $i$ at layer $l-1$. $\frac{\partial{E}}{\partial{t_j^{(l)}}}$ is the error transfer function which can be calculated recursively following
\begin{equation}
\frac{\partial{E}}{\partial{t_j^{(l-1)}}} = \frac{\partial{o_i^{(l-1)}}}{\partial{t_i^{(l-1)}}} \times \sum_{j=1}^{L}\frac{\partial{E}}{\partial{t_j^{(l)}}}\times \frac{\partial{t_j^{(l)}}}{\partial{o_i^{(l-1)}}}
\end{equation}
where $\frac{\partial{t_j^{(l)}}}{\partial{o_i^{(l-1)}}}=w_{i,j}^{(l)}$. For the output tanh layer we have $\frac{\partial{o_i^{(L)}}}{\partial{t_i^{(L)}}}=\text{sech}^2(t_i^{(L)})$. Once we have the gradients of error function w.r.t to the weight parameters, we can fine-tune the network with error back-propagation.

\subsection{Corrupted data}

\begin{figure}[t]
\begin{center}
\subfigure[Original]{\includegraphics[height=5.3cm]{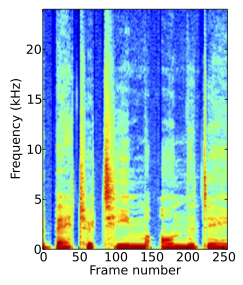}}
\hspace{-3mm}
\subfigure[Masked]{\includegraphics[height=5.3cm]{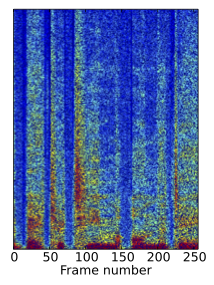}}
\end{center}
\vspace{-0.5cm}
\caption{These figures shows parts of original and masked spectra. In the right figure black points indicated masked regions.}
\label{fig:mask}
\vspace{-3mm}
\end{figure}
We used a masking technique reported in \cite{ref:Vincent08} to corrupt the training data for the denoising auto-encoder.
This technique independently and randomly set the values of the training data in different dimensions to zero following a Bernoulli distribution.
Figure \ref{fig:mask} shows an example of original and masked spectra.
In this figure, black points indicate masked regions.
% \textbf{Zhenzhou: Please mention how you create corrupted date $\tilde{X}$. Please also add a figure before and after you mask the spectrum}

\section{Evaluation}

This section shows experimental results. We have evaluated the proposed auto-encoder method in the context of analysis-by-synthesis condition and text-to-speech conditions. In the text-to-speech experiments, the synthetic voices using the proposed acoustic features were modeled using two state-of-the-art speech synthesis systems: HMM and DNN.

\subsection{Dataset}

The dataset we use consists of 4569 short audio waveforms uttered by a professional English female speaker and each waveform is around 5 seconds long. For each waveform, we first extract its frequency spectra using STRAIGHT vocoder with 2049 FFT points. We then extract the low dimensional feature from each 2049-dim STRAIGHT spectrum using autoencoder. All data was sampled at $48$ kHz. For comparison of the proposed method, we extracted mel-cepstral coefficients that use the same dimensions as that of auto-encoder. All other acoustic features such as log F0 and $25$ aperiodicity band energies are the same for all the systems.

\subsection{Configurations of the deep denoising auto-encoder}

\begin{figure}[t]
  \begin{center}
  \includegraphics[width=8.0cm]{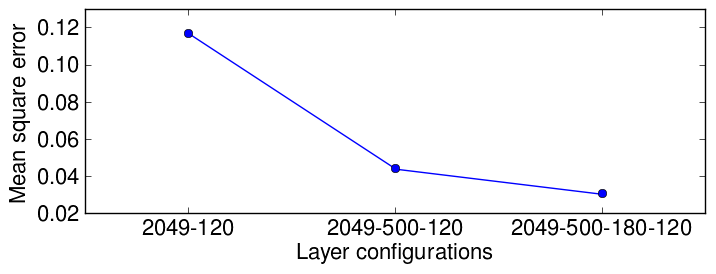}
  \end{center}
\vspace{-0.5cm}
\caption{Reconstruction mean square errors for auto-encoders of different architectures but same bottleneck dimension.}
\label{fig:mse}
\vspace{-3mm}
\end{figure}
Figure \ref{fig:mse} shows the reconstruction mean square errors of auto-encoders trained on raw frequency-warped spectrum with different number of hidden layers. It shows that the error decreases with more hidden layers, and that deep auto-encoder is better than shallow auto-encoder with the same bottleneck dimension. For the results in the rest of paper, we use architecture of the auto-encoder as 2049-500-180-120 for producing the 120-dim acoustic features, tanh units for all the layers and the inputs are 2049-dim Bark-scale-based frequency-warped spectrum, which are preprocessed with global contrast normalization. The hyperparameters used for the layer-by-layer pre-training are searched randomly and the set of values that produce the best results are selected. Table \ref{tb:hyper} shows the hyperparameters for the auto-encoders used in the experiments.
\begin{table}[t]
\caption{The table lists down the hyperparameters used for training each model. lr: learning rate, m: momentum, b: batch size, s: numpy random variable weight initialization seed \cite{ref:Ilya13}, d: masking probability of each input dimension \cite{ref:Vincent08}.}
\label{tb:hyper}
\vspace{-1em}
\begin{center}
\begin{tabular}{l|c|c|c|c|c|c|c}
\cline{2-7}
% & & \multicolumn{4}{ l| }{Primes} \\ \cline{3-6}
& layer dim& lr & m & b & s & d &\\ \cline{1-7}
%\multicolumn{1}{ |l  }{\multirow{4}{*}{\begin{minipage}{0.5in}DAE-120\end{minipage}} } &
%\multicolumn{1}{ |c| }{2049-500} & 0.01 & 0.01 & 100 & 3158 &   0.5  \\ \cline{2-7}
%\multicolumn{1}{ |c  }{}                        &
%\multicolumn{1}{ |c| }{500-180} & 0.01 & 0.5 & 100 & 1743 &  0.3   \\ \cline{2-7}
%\multicolumn{1}{ |c  }{}                        &
%\multicolumn{1}{ |c| }{180-120} & 0.1 & 0.01 & 50 & 3188 & 0.5    \\ \cline{2-7}
%\multicolumn{1}{ |c  }{}                        &
%\multicolumn{1}{ |c| }{Finetune} & 0.001 & 0.9 & 50 & 9548 & N.A    \\ \cline{1-7}
\multicolumn{1}{ |c  }{\multirow{4}{*}{\begin{minipage}{0.5in}Deep auto-encoder\end{minipage}} } &
\multicolumn{1}{ |c| }{2049-500} & 0.001 & 0.9 & 200 & 8963 &  N.A   \\ \cline{2-7}
\multicolumn{1}{ |c  }{}                        &
\multicolumn{1}{ |c| }{500-180} & 0.01 & 0.5 & 50 & 1902 & N.A    \\ \cline{2-7}
\multicolumn{1}{ |c  }{}                        &
\multicolumn{1}{ |c| }{180-120} & 0.01 & 0.9 & 50 & 6555 & N.A    \\ \cline{2-7}
\multicolumn{1}{ |c  }{}                        &
\multicolumn{1}{ |c| }{Finetune} & 0.01 & 0.5 & 150 & 9781 & N.A     \\ \cline{1-7}
\multicolumn{1}{ |c  }{\multirow{4}{*}{\begin{minipage}{0.5in}Deep denoising auto-encoder\end{minipage}} } &
\multicolumn{1}{ |c| }{2049-500} & 0.01 & 0.1 & 150 & 5252 & 0.1    \\ \cline{2-7}
\multicolumn{1}{ |c  }{}                        &
\multicolumn{1}{ |c| }{500-180} & 0.01 & 0.5 & 150 & 7514 & 0.1    \\ \cline{2-7}
\multicolumn{1}{ |c  }{}                        &
\multicolumn{1}{ |c| }{180-120} & 0.01 & 0.9 & 100 & 594 & 0.5    \\ \cline{2-7}
\multicolumn{1}{ |c  }{}                        &
\multicolumn{1}{ |c| }{Finetune} & 0.001 & 0.9 & 100 & 2208 & N.A    \\ \cline{1-7}
\end{tabular}
\end{center}
\vspace{-5mm}
\end{table}

% The warp frequency spectral that is used to train the autoencoders for the text-to-speech experiments is scaled to [-0.5, 0.5]. During the layer by layer pretrainig of the deep autoencoder 2049-500-180, the learning rate r, momentum m, batch size z, weight initialization seed s for pretraining 2049-500 is (), also (r, m, z, s) for pretraining 500-180 is (), and (r, m, z, s) for pretraining 180-120 is ().
% \textbf{Zhenzhou: Please describe pre-processing, number of layers, dimension of each layer, step size, activate function, dropout, percentages of masked data etc}
% \textbf{Zhenzhou: Please add a figure for showing the effects of using many layers for DAE. Horizontal axis is number of layers, Vertical axis is MSE.}

\subsection{Analysis-by-synthesis experimental results}

First we report the analysis-by-synthesis experimental results. For this evaluation, we have divided the above database into three subsets, that is, training, validation and test. The training subset was used as training data for building the auto-encoder, the validation subset was used as a stopping criteria during training to prevent over-fitting, and the test subset was used for measuring log-spectral distortion and listening test.

\begin{figure*}[t]
\begin{center}
\subfigure[Original]{\includegraphics[height=5.3cm]{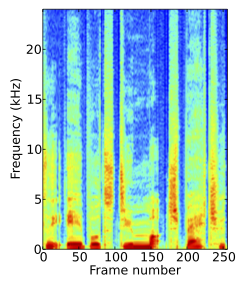}}
\hspace{-3mm}
\subfigure[mel-cepstrum]{\includegraphics[height=5.3cm]{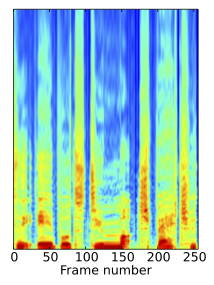}}
\hspace{-3mm}
% \subfigure[AE-120]{\includegraphics[height=5.3cm]{./images/analysis-synthesis_reconstructed-spec/DeepAutoEncoder-120.png}}
% \subfigure[DAE-120]{\includegraphics[height=5.3cm]{./images/analysis-synthesis_reconstructed-spec/DenoisingDeepAutoEncoder-120.png}}
\subfigure[deep auto-encoder]{\includegraphics[height=5.3cm]{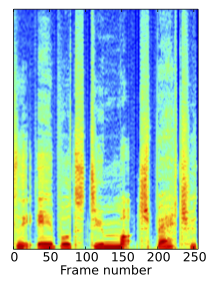}}
\hspace{-3mm}
\subfigure[deep denoising auto-encoder]{\includegraphics[height=5.3cm]{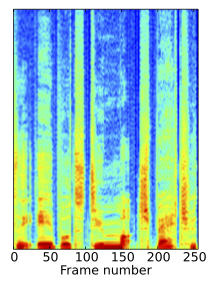}}
\end{center}
\vspace{-0.5cm}
\caption{Original and reconstructed spectra using each technique.}
\label{spectrum}
\end{figure*}

\begin{figure}[t]
  \begin{center}
  \includegraphics[width=6.5cm]{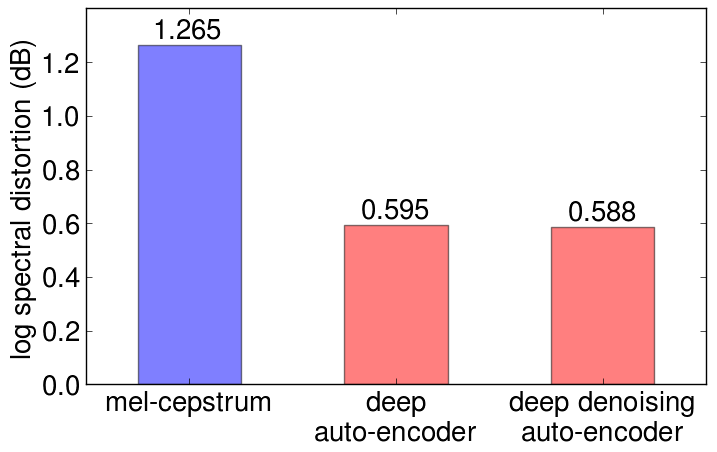}
  \end{center}
  \vspace{-0.5cm}
  \caption{log spectral distortion between the original and reconstructed spectra.}
  \label{fig:log_spectral_distortion}
  \vspace{-3mm}
\end{figure}

\begin{figure}[t]
  \begin{center}
  \includegraphics[width=8.5cm]{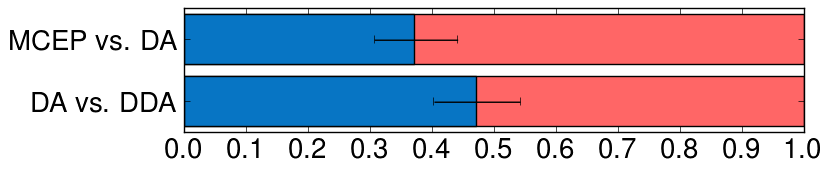}
  \end{center}
\vspace{-0.5cm}
\caption{Results of preference tests using analysis-by-synthesis speech samples. In this figure, MCEP, DA and DDA refer to mel-cepstrum analysis, deep auto-encoder and deep denoising auto-encoder respectively.}
\label{fig:results_analysis-by-synthesis}
\vspace{-5mm}
\end{figure}

Figure \ref{spectrum} shows the original and reconstructed spectra using each technique (mel-cepstral analysis, deep auto-encoder, deep denoising auto-encoder). We can clearly see that the deep auto-encoders reconstruct high-frequency parts more precisely than mel-cepstral analysis.
Figure \ref{fig:log_spectral_distortion} shows log spectral distortion between the original spectra and reconstructed spectra, calculated on the test subset. We can observe that the deep auto-encoder has reduced the distortion significantly compared to the mel-cepstral analysis and denoising version further reduced the distortion.
Figure \ref{fig:results_analysis-by-synthesis} shows subjective preference scores of these methods. The number of listeners that performed this test were seven. They have participated in two preference tests. In the first preference test, they were asked to compare deep auto-encoder (DA) with mel-cepstral analysis (MCEP). In the second preference test, they were asked to compare deep auto-encoder with deep denoising autoencoder (DDA). From the figure, we can see that deep auto-encoder based speech samples sound more natural than mel-cesptral analysis based speech samples. Deep denoising auto-encoder reduced the distortion, however, perceptual difference between clean and denoising auto-encoder is not statistically significant.
%
%\begin{table}[t]
%\caption{Results of preference tests using analysis-by-synthesis speech %samples.}
%\label{tb:results_analysis-by-synthesis}
%\begin{center}
%\begin{tabular}{|c|c|}
%\hline
%MCEP : DA  & $37.1\%$ : $62.9\%$ ($p < 0.05$) \\\hline
%DA   : DDA & $47.2\%$ : $52.8\%$ ($p = 0.50$) \\\hline
%\end{tabular}
%\end{center}
%\end{table}
%

\subsection{Text-to-speech experimental results}

Next we report the text-to-speech experimental results. For the HMM-based speech synthesis, we have used a hidden semi-Markov model and the observation vectors for the spectral and excitation parameters contained static, delta and delta-delta values, with one stream for the spectrum, three streams for F$0$ and one for the band-limited aperiodicity. The context-dependnet labels are built using the pronunciation lexicon Combilex \cite{ref:Richmond10}. For the DNN-based speech synthesis, we have trained a five-hidden-layer DNN for mapping between linguistic contexts and auto-encoder-based or mel-cepstral acoustic features. The number of units in each of the hidden layers was set to 512. Random initialisation was used in a similar way to \cite{ref:Zen13}.
%
%\begin{table}[t]
%\caption{Preference scores ($\%$) between text-to-speech samples from MCEP and AE for HMM and DNN-based speech %synthesis systems.}
%\label{tb:results_tts}
%\begin{center}
%\begin{tabular}{|c|c|}
%\hline
%MCEP-HMM : DA-HMM & $40.5\%$ : $59.5\%$ ($p < 0.05$) \\\hline
%MCEP-DNN : DA-DNN & $21.0\%$ : $79.0\%$ ($p < 0.05$) \\\hline
%\end{tabular}
%\end{center}
%\end{table}
%
\begin{figure}[t]
  \begin{center}
  \includegraphics[width=8.5cm]{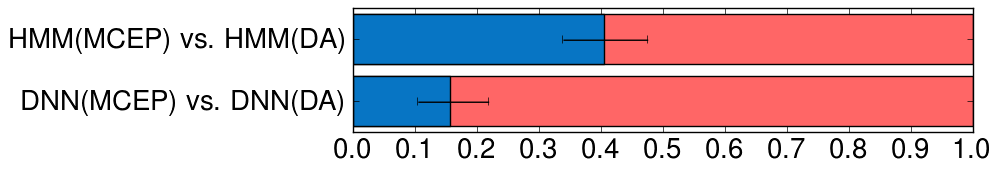}
  \end{center}
\vspace{-0.5cm}
\caption{Results of preference tests using text-to-speech samples. In this figure, MCEP and DA refer to mel-cepstrum analysis and deep auto-encoder for the acoustic feature extraction, and HMM and DNN are the acoustic models.}
\label{fig:results_tts}
\end{figure}
Figure \ref{fig:results_tts} shows subjective preference scores where we have compared the proposed auto-encoder feature with the conventional mel-cepstral feature in each of the HMM-based speech synthesis and the DNN-based speech synthesis systems. Listeners are the same as those for Figure \ref{fig:results_analysis-by-synthesis}. We can see that synthetic speech using the proposed feature sound more natural than the conventional mel-cepstral features in both the synthesis methods. The proposed feature seems to suit the DNN-based speech synthesis better, but, this requires further investigation.

\section{Conclusions}

In this paper we have proposed the deep denoising auto-encoder technique to extract better acoustic features for speech synthesis. We have compared the new stochastic feature extractor with the conventional mel-cepstral analysis in the analysis-by-synthesis and text-to-speech experiments and have confirmed that the proposed method can increase the quality of synthetic speech in both the conditions.

Our future work includes the improvement of the deep denoising auto-encoder. In this paper, we have used the simplest noise, i.e. masking and the improvement was observed only from objective evaluation. We shall use or design different types of noises to improve the deep denoising auto-encoder for speech synthesis further.

% Below is an example of how to insert images. Delete the ``\vspace'' line,
% uncomment the preceding line ``\centerline...'' and replace ``imageX.ps''
% with a suitable PostScript file name.
% -------------------------------------------------------------------------

\vfill
\pagebreak

\nocite{ref:Bergstra12}
\nocite{ref:Xue14}
\nocite{ref:Remelhart86}

%\bibliographystyle{IEEEbib}
%\bibliography{output.bbl}
%\bibliography{references}
%\addbibresource{references.bib}

\end{document}